\documentclass{iopart}
\usepackage{graphicx}

\begin{document}

\title{How can a 22-pole ion trap exhibit 10 local minima in the effective
  potential?}

\author{R. Otto$^1$, P. Hlavenka$^1$, S. Trippel$^1$,
  J. Mikosch$^1$\footnote{present address: National Research Council Canada,
    100 Sussex Drive, Ottawa,ON K1A 0R6, Canada}, K. Singer$^2$,
  M. Weidem{\"u}ller$^1$\footnote{present address: Physikalisches Institut,
    Universit{\"a}t Heidelberg, Philosophenweg 12, 69120 Heidelberg, Germany},
  R. Wester$^1$}

\address{$^1$ Physikalisches Institut, Universit{\"a}t Freiburg,
Hermann-Herder-Stra{\ss}e 3, 79104 Freiburg, Germany}

\address{$^2$ Institut f{\"u}r Quanteninformationsverarbeitung, Universit{\"a}t
  Ulm, Albert-Einstein-Allee 11, 89081 Ulm, Germany}

\ead{roland.wester@physik.uni-freiburg.de}

\date{\today}

\begin{abstract}
\noindent
The column density distribution of trapped OH$^-$ ions in a 22-pole ion trap
is measured for different trap parameters. The density is obtained from
position-dependent photodetachment rate measurements. Overall, agreement is
found with the effective potential of an ideal 22-pole. However, in addition
we observe 10 distinct minima in the trapping potential, which indicate a
breaking of the 22-fold symmetry. Numerical simulations show that a
displacement of a subset of the radiofrequency electrodes can serve as an
explanation for this symmetry breaking.
\end{abstract}


\maketitle


\section{Introduction}
Multipole radiofrequency ion traps \cite{gerlich1992:adv}, in particular the
22-pole ion trap \cite{gerlich1995:ps}, are versatile devices used in laser
spectroscopy
\cite{Schlemmer1999Laser,schlemmer2002Laser,mikosch2004:jcp,asvany2005:sci,Glosik06,boyarkin2006:jacs,dzhonson2006:jms,kreckel2008:jcp}
and investigations of chemical reaction processes
\cite{gerlich1995:ps,asvany2004:cp,asvany2004:apj,otto2008:prl} of atomic and
molecular ions. High order multipole traps offer a large field free region in
the trap center, and therefore provide a reduced interaction time of the ions
with the oscillating electric field compared to a quadrupole trap
\cite{wester2009:jpb}. Buffer gas cooling down to cryogenic temperatures of
the translational \cite{Schlemmer1999Laser,Glosik06}, rotational
\cite{Schlemmer1999Laser,dzhonson2006:jms}, and vibrational degrees of freedom
\cite{boyarkin2006:jacs} of trapped molecular ions has been demonstrated. This
enables applications with all the advantages of low temperature experiments,
such as a reduced Doppler width in spectroscopy studies and a well defined
population of internal states.

Stable confinement of a single ion in the oscillating quadrupole field of a
Paul trap is precisely predicted, because the Mathieu equations of motion can
be solved analytically. This is different for oscillating high order multipole
fields, where the equations of motion have no analytical solution. However,
the movement of the ions in a fast oscillating rf-field justifies the
assumption of an effective trapping potential, based on the separation of the
ion motion into a smooth drift and a rapid oscillation, called micromotion
\cite{dehmelt1967:adv}. This effective potential can be expressed as
\begin{equation}
\label{effective_potential}
V^*(\vec r) = \frac{1}{4} \frac {(q \vec{E}(\vec r) )^2} {m\Omega^2} + q
\Phi_0
\end{equation}
where $\vec{E}(\vec r)$ denotes the electric field at the point $\vec r$, $m$
is the mass of a test particle with charge $q$ in an electric field
oscillating on frequency $\Omega$ and $\Phi_0$ is a non-oscillating dc
potential.  For an ideal cylindrical multipole of the order $n$ this effective
potential can be expressed as
\begin{equation}
\label{multipole_effective_potential}
V^*(r) = \frac{1}{4} \ \frac{n^2 (q V_0)^2}{m\Omega^2r_0^2} \
\left(\frac{r}{r_0}\right)^{2n-2}
\end{equation}
where $V_0$ denotes the amplitude of the oscillating rf-field. For a high
order multipole field ($n$\,=\,11 for the 22-pole trap) this creates an almost
box-like trapping volume with steep walls and a large field free region in the
center. This is important to suppress rf heating when cooling ions in neutral
buffer gas \cite{wester2009:jpb}.

The number of trapped ions that can be excited with radiation depends on the
local density of ions in the interaction volume with the light field. In
experiments with trapped Ba$^+$ ions in an octupole trap the density has been
imaged by spatially resolving the fluorescence signal
\cite{schubert1989:josa,walz94}. In this article we report on a method based
on the photodetachment of trapped anions that allows us to image the density
distribution of the trapped ions. In this way we measure the experimental
trapping potential of a 22-pole trap with unprecedented accuracy and study its
dependence on the radiofrequency field. In the next section the
photodetachment tomography is explained, followed by a presentation of
measured density distributions at different temperatures and rf amplitudes. We
find ten unexpected maxima in the density distribution and analyze their
origin with the help of numerical simulations in section \ref{model}.

\section{Experimental procedure} 
The experimental radiofrequency ion trap setup has been described in detail in
Ref.  \cite{mikosch2008:pra}. Its central element is a 22-pole trap
\cite{gerlich1995:ps} with good optical access along the trap axis. In the
present tomography experiments, we have used an ensemble of stored OH$^-$
anions cooled to a translational temperature of either 170 or 300\ K. The use
of negative ions allows us to perform photodetachment measurements, were the
anions are depleted by a laser beam propagating parallel to the symmetry axis
of the trap \cite{trippel2006:prl,hlavenka2009:acc}. The photodetachment
process
\begin{equation}
\rm{OH}^- + h\nu \rightarrow \rm{OH} + e^-
\end{equation}
with $h\nu > 1.83$\,eV yields a depletion rate proportional to the
photodetachment cross section and to the overlap of the ion column density
with the photon flux. By scanning the position of the laser beam and measuring
the depletion rate at each position the relative ion column density
distribution is obtained. For small ion densities, where the Coulomb
interaction between the ions can be neglected, the depletion rate at the
transverse position $(x,y)$ can be expressed as
\begin{equation} 
k(x,y) =  \sigma_{pd} F_L \rho(x,y).
\end{equation} 
It depends only on the total photon flux $F_L$, the photodetachment cross
section $\sigma_{pd}$ and the single particle column density $\rho(x,y)$,
reflecting the spatial overlap of the laser beam with the ion distribution
\cite{trippel2006:prl}. As the latter is normalized to unity, we can write it
as
\begin{equation} 
\rho(x,y) =  \frac{k(x,y)}{\int_S{k(x,y)}dS}
\end{equation} 
Thus, a full two-dimensional tomography scan of the trapped ions can be used
to map the entire ion column density in the trap. If $k(x,y)$ is measured for
the whole ion distribution, also the absolute photodetachment cross section
can be obtained \cite{hlavenka2009:acc}.

Ions are produced in a pulsed supersonic expansion of a suitable precursor
gas, crossed by a 1\,keV electron beam. For the OH$^-$ production we use a
mixture of Ar/NH$_3$/H$_2$O (88\%, 10\%, 2\%) to create NH$_2^-$. A rapid
chemical conversion by water forms the OH$^-$ anions in the source. A bunch of
$\sim$500 mass-selected ions is loaded into the trap, which is enclosed by a
copper housing that is temperature-variable between 8 and 300\,K. A typical He
buffer gas density of $1\times10^{14}\,$cm$^{-3}$, employed for all
temperatures, is enhanced by a buffer gas pulse during the ion injection. The
trap is operated with $\Omega = 2 \pi \times 5$\,MHz radiofrequency and
different rf amplitudes. Along the axial direction ions are confined by end
electrodes biased to -2\,V. To allow good thermalization of the ions, a
storage period of 200\,ms is inserted before the laser beam is switched
on. After a given storage and laser interaction time the current signal,
proportional to the number of ions that survived the interaction with the
photodetachment laser, is detected with a microchannel plate.

The two-dimensional tomography scans are performed under the same experimental
conditions as in Ref.\ \cite{hlavenka2009:acc}. A free-running continuous wave
diode laser at 661.9 nm (Mitsubishi ML1J27, 100mW, spectral width 0.7 nm FWHM)
has been employed. It is imaged into the trap and scanned in the $(x,y)$
plane, perpendicular to the symmetry axis of the trap, by moving the imaging
lense on a mesh with 0.25\,mm point spacing with a computer-controlled
two-dimensional translation stage. The mesh spacing is comparable to the
$1/e^2$ radius of the laser in the trap of 350\,$\mu$m. For each laser
position the photodetachment depletion rate is obtained in a storage time
interval of 0.2-2\,s by fitting an exponential decay to the ion current
signal, reduced by typically one percent due to the background loss rate. The
data are averaged over typically 4-8 scans. Within each scan the mesh points
are accessed in random order to avoid systematic drifts.

\section{Tomography of the trapping potential}
\begin{figure}[b]
  \begin{center}
    \includegraphics[angle=0,trim=5 0 0 15, width=\columnwidth]{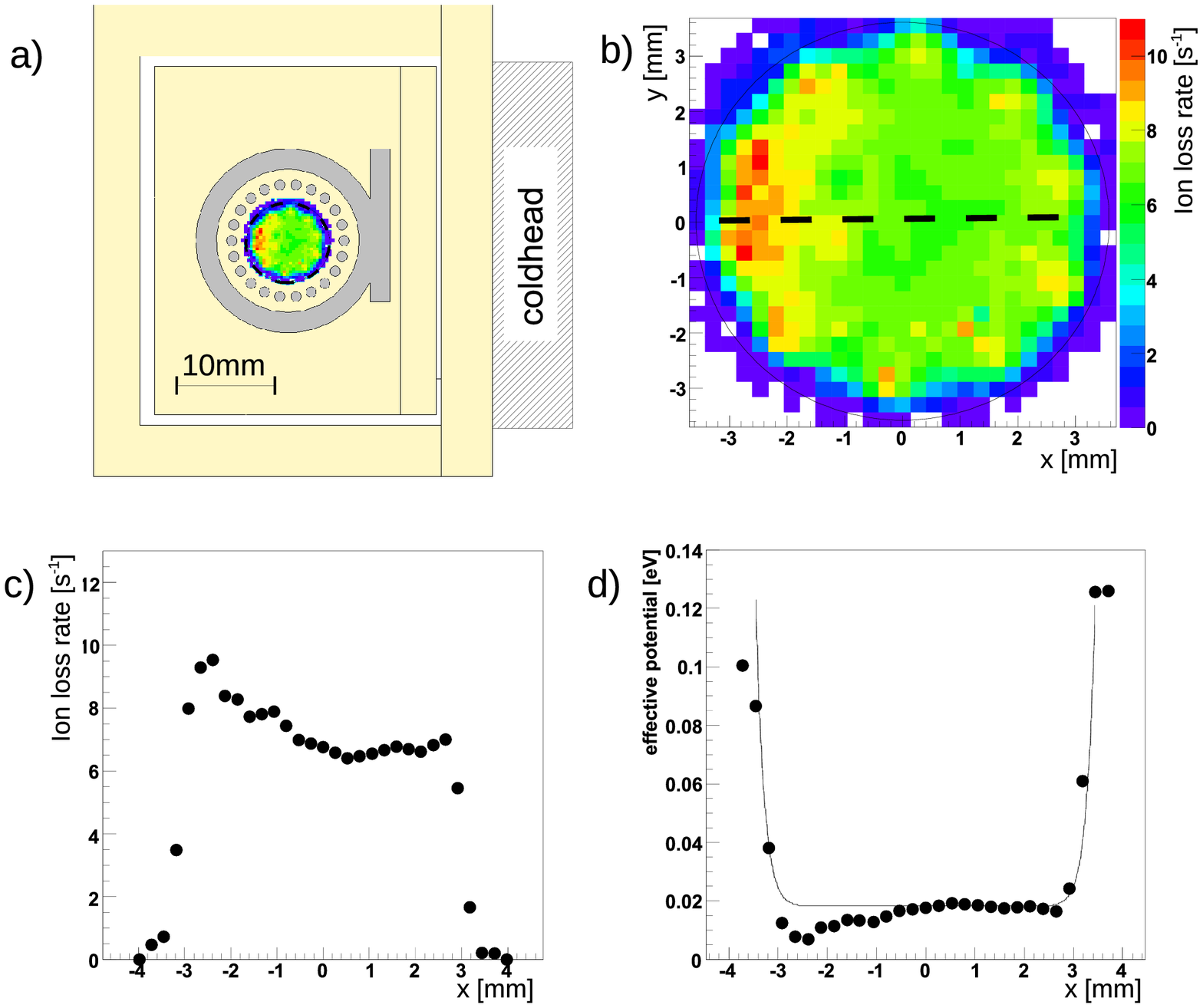}
    \caption{(Color online) a) Measured density distribution of trapped OH$^-$
      ions at 300\,K buffer gas temperature with the rf amplitude set to
      160\,V. The sketched geometry shows the layout of the ion trap, viewed
      along its symmetry axis. It includes the copper housing, the 22 rf
      electrodes (end-on), a surrounding shaping electrode, and the end
      electrodes. b) Zoom into the measured ion density distribution, each
      pixel represents an individual decay rate measurement. c)
      One-dimensional cut through the density distribution along the
      horizontal axis. d) Effective potential derived from the density
      distribution by assuming a Boltzmanm distribution of the trapped ions at
      300\,K.}
\label{fig1}
 \end{center}
\end{figure}

Fig. \ref{fig1}a shows a tomography scan of OH$^-$ anions in the 22-pole trap
at 300\,K with the rf amplitude set to 160\,V. The figure also contains the
sketched arrangement of the trap's copper housing mounted on the coldhead of
the cryostat, the position of the 22 rf electrodes and the axial end
electrodes. Fig. \ref{fig1}b is a zoom of the scan which more clearly shows
the measured ion density distribution. Every pixel of the histogram here
represents a fitted photodetachment depletion rate $k(x,y)$ (see previous
section) and is proportional to the single-particle probability density
$\rho(x,y)$ along a column parallel to the $z$-axis. As can be seen the ion
distribution as a first approximation can be considered circularly symmetric
and constant in the center region of the trap, whereas it drops to zero when
the ions reach the outer regions of the trapping volume. This distribution
directly visualizes the overall storage properties of a 22-pole ion trap with
a flat potential in the center and steep walls. Note that the ion density
drops to zero already for smaller radii than the end electrode (solid line in
Fig. \ref{fig1}b), indicating that clipping of the laser at the end electrode
is not affecting the measured density distribution. For smaller rf amplitudes
the ion density distribution would extend to larger radii and could not be
fully probed by the photodetachment tomography. For this reason we have
restricted ourselves to large enough rf amplitudes in this study.
 
In Fig. \ref{fig1}c a horizontal cut through the ion distribution is
shown. While in the center the distribution is relatively uniform, the
population is locally enhanced by up to 40\% near the edge of the ion
distribution. Such a behavior has already been observed in previous
measurements \cite{trippel2006:prl}. To study this here in more detail, the
effective potential $V(x,y)$ is extracted from the local ion density
$\rho(x,y)$ assuming a Boltzmann distribution for the ions in the trap
\begin{equation}
\rho(x,y) = \frac{1}{Z} \exp(-V(x,y)/k_{\rm B}T),
\label{density:eq}
\end{equation}
where $T$ is the absolute temperature, $k_{\rm B}$ is Boltzmann's constant,
and $Z$ is the partition function. Since only the column density is measured,
the resulting potential $V(x,y)$ is an average over the $z$-direction. Fig.\
\ref{fig1}d shows a cut through the obtained effective potential for the
distribution of Fig.\ \ref{fig1}c. Overall this potential compares well with
the calculated potential of an ideal 22-pole potential (solid line), obtained
without any free parameters from Eq.\
(\ref{multipole_effective_potential}). Closer inspection reveals interesting
features in the potential that deviate from the ideal multipole. While the
potential is relatively flat in the center, it shows a distinct minimum of
about 12\,meV near the left edge of the ion distribution and a weaker minimum
of about 5\,meV near its right edge. It can be excluded that this change of
the distribution is caused by space charge effects, because the experiments
are performed with only a few hundred ions in a trap volume of about
1\,cm$^3$.

\begin{figure}[b]
  \begin{center}
    \includegraphics[angle=0,trim=5 0 0 15, width=\columnwidth]{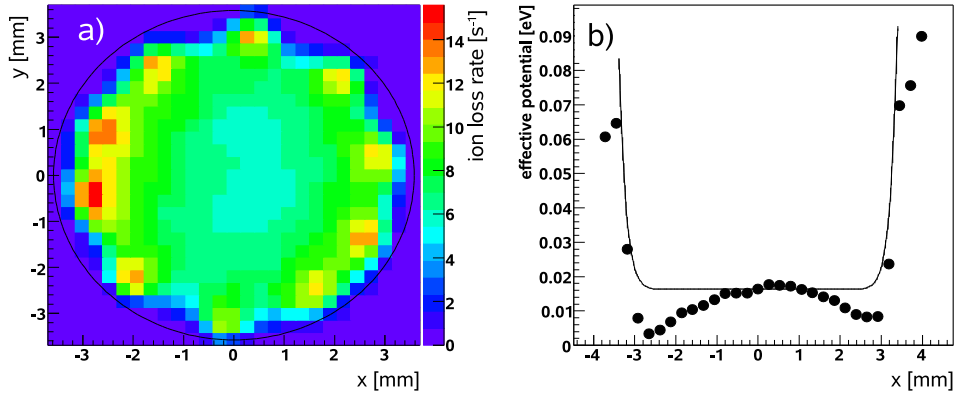}
    \caption{(Color online) a) Ion density distribution at 170\,K with an rf
amplitude of 160\,V. It shows a substructure of ten clearly distinct maxima.
b) Cut through the effective potential, derived from the the density
distribution. Overall, the potential is in accordance with the effective
potential of an ideal 22-pole (solid line).}
\label{fig2}
  \end{center}
\end{figure}

The same features of the effective potential are observed in measurements at a
lower trap temperature. Fig. \ref{fig2}a shows a tomography scan at 170\,K and
the same rf amplitude as above. The ion distribution again looks circular
symmetric with a distinct cutoff when the ions reach the steep walls of the
trapping potential. A horizontal cut through the effective potential, obtained
in the same fashion as Fig. \ref{fig1}d, is shown in Fig. \ref{fig2}b. The
same minima as for 300\,K are observed in the effective potential. At this
lower temperature the two minima are better resolved and appear similar in
depth on the left side and slightly deeper on the right side of the potential
as compared to the 300\,K tomography.

\begin{figure}[b]
  \begin{center}
    \includegraphics[angle=0,trim=5 0 0 15, width=\columnwidth]{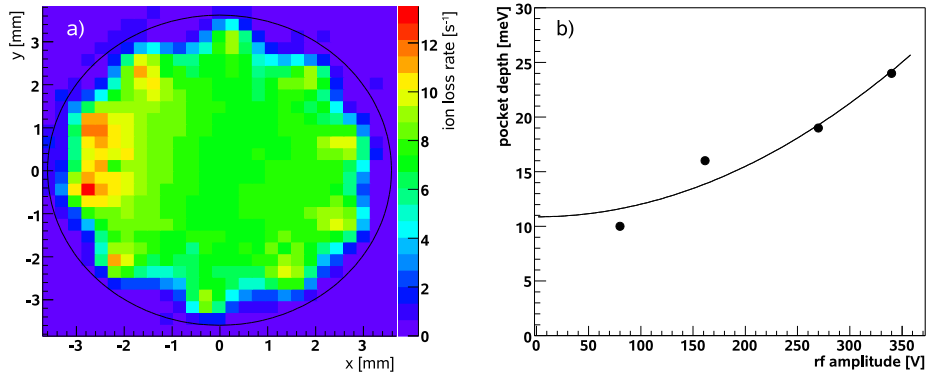}
    \caption{(Color online) a) Ion density distribution at 300\,K with an
rf-amplitude of 270\,V. The histogram shows ten distinct maxima, which
correspond to minima in the effective potential. b) The depth of the deepest
minimum shows a strong increase as a function of the applied rf-amplitude. The
solid line shows a polynomial fit with only a constant and a quadratic term.}
\label{fig3}
  \end{center}
\end{figure}

Further substructure becomes visible in the 170\,K density distribution. Ten
clearly separated maxima in the density distribution appear almost equally
spaced in angle at a radial position of about 3\,mm. According to Eq.\
(\ref{density:eq}) they correspond to ten localized minima in the trapping
potential at this radius with a typical depth of 10\,meV. These minima have
not been significant in the 300\,K ion distribution at 160\,V rf
amplitude. They become visible, however, for larger amplitudes. Fig.\
\ref{fig3}a shows a 300\,K ion distribution for an rf amplitude of 270\,V. It
reveals the same ten density maxima and respective potential minima that could
only be resolved at lower temperature at 160\,V.

We have studied the dependence of the depth of the ten potential minima on the
rf amplitude at 300\,K. The depth of the deepest minimum is plotted in
Fig. \ref{fig3}b. Since the effective potential is expected to depend
quadratically on the rf amplitude, a fit with only a constant and a quadratic
term is applied to the data (solid line in Fig.\ \ref{fig3}b). It yields an
rf-independent offset of about 11\,meV, which is attributed to the static
potential of the end electrodes of the ion trap. These end electrodes produce
a radially repulsive potential inside the trap, as discussed in Ref.\
\cite{trippel2006:prl,asvany2009:ijm}. It compares well with simulations, as
shown in the next section. The ten ``pockets'' in the potential, however,
reveal a more complex deviation from the ideal multipole description of Eq.\
(\ref{multipole_effective_potential}).  An explanation for them will also be
discussed in the next section.

\section{Modeling trapping potentials of realistic multipoles
\label {model}}
The effective potential of a $2n$-pole has a $2n$-fold rotational symmetry,
when averaging over one radiofrequency period. The appearance of the ten
observed potential minima is therefore a clear indication for a breaking of
the ideal symmetry. To investigate this effect further, the effective
potential of the employed 22-pole trap has been modeled using a numerical
simulation package based on a fast multipole solver
\cite{greengard1988,nabors1994:jsc} for the boundary element problem in
combination with accurate field evaluation in free space. With this method the
electric field $E(r)$ can be calculated at any location inside the trap. It is
converted into the effective trapping potential using
Eq. (\ref{effective_potential}). We have verified that the simulation of the
trapping potential of an ideal 22-pole structure reproduces the effective
potential of Eq.\ (\ref{multipole_effective_potential}) on the numerical level
of accuracy.

\begin{figure}[b]
  \begin{center}
    \includegraphics[angle=0,trim=5 0 0 15, width=\columnwidth]{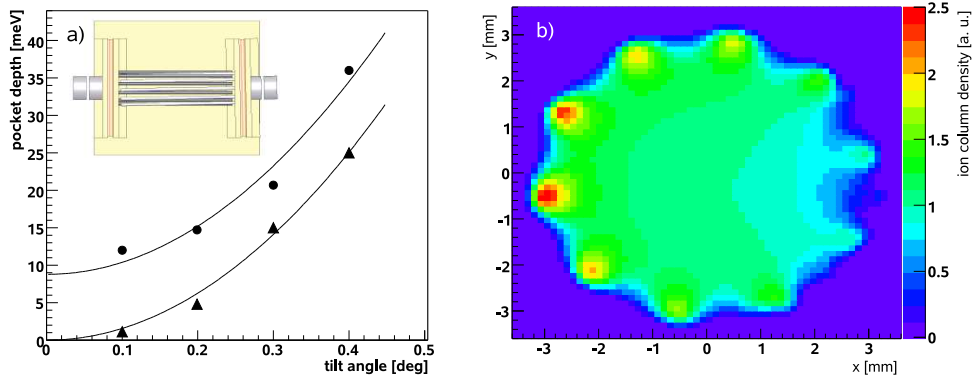}
    \caption{(Color online) a) Depth of the deepest potential minimum as a
function of the tilt angle of the rf electrodes with a potential applied to
the end electrodes (upper points) and without any potential on the end
electrodes (lower points). The solid lines show quadratic fits without a
linear term.  The inset shows the geometry of the 22-pole ion trap, viewed in
the direction onto the coldhead (see Fig.\ \ref{fig1}a). The right wall of the
trap together with the 11 implanted rf electrodes has been tilted by an angle
of $1.0^\circ$. b) Calculated ion density distribution for 300\,K in the
effective potential of 270\,V rf amplitude and -2\,V static end electrode
potential. A tilt angle of 0.15$^\circ$ is chosen, which leads to a good
agreement with the measured density distribution shown in Fig.\ \ref{fig3}a.}
\label{fig4}
   \end{center}
\end{figure}

Different assumptions have been tested as origin of the observed ten potential
minima, such as the influence of the shape and position of the end electrodes
and of the copper housing around the trap electrodes, without showing a
measurable effect on the potential. This suggests that imperfections of the
trap geometry itself may be responsible. To simulate these imperfections a
breaking of the ideal symmetry is introduced by displacing one half of the 22
radiofrequency electrodes by a small angle (see inset in Fig.\
\ref{fig4}a). Such a small tilt of one set of electrodes against the others
occurs to be the most likely displacement during the trap assembly. Upon
tilting one set of electrodes by only a few tenths of a degree the calculated
effective potential of a 22-pole trap at 160\,V rf amplitude and -2\,V on the
end electrodes immediately shows ten potential minima.

In Fig.\ \ref{fig4}a the dependence of the maximum pocket depth on the tilt
angle, as obtained from a series of simulations, is plotted. These simulations
have been carried out for 160\,V rf amplitude. Here, an imperfection in the
parallelity of only $0.2^\circ$ causes a pocket depth of 5\,meV. The pocket
depth is calculated at each angle with and without a potential of -2\,V
applied to the static end electrodes. The end electrode voltage produces an
overall quadrupole potential that pushes the ion ensemble towards larger radii
in the trap in addition to the tilt-induced pockets. Both data sets with and
without end electrode potential are described by the same quadratic
increase. For the simulations with end electrode potential a constant offset
of about 9\,meV is obtained, in fair agreement with the experiment value of
about 11\,meV.

From the measured depths of the potential minima (Fig.\ \ref{fig3}b) a value
of between 3 and 5\,meV is extracted for an rf amplitude of about 160\,V,
after subtracting the influence of the static end electrode (see Fig.\
\ref{fig3}b). Such an rf-induced pocket depth is obtained in the simulation
for a tilt angle of between $0.15^\circ$ and $0.2^\circ$ (see Fig.\
\ref{fig4}a). Fig.\ \ref{fig4}b shows a simulated density distribution for a
tilt angle of $0.15^\circ$. Following Eq.\ (\ref{density:eq}), the simulation
has been performed for OH$^-$ ions that are stored at 300\,K in the 22-pole
trap with 270\,V rf amplitude and -2\,V potential on the end electrodes. This
simulated density distribution agrees well with the measured distribution of
Fig.\ \ref{fig3}a, which has been obtained with the same trap parameters. A
larger tilt angle was found to already overestimate the ten potential
minima. Note that it is preferable to compare graphs of the density
distributions of simulation and experiment instead of effective potentials,
because the experimental potential is obtained by a logarithm of the density
distribution which suppresses the fine details in the images.

When the 22-pole trap was assembled the strong influence of small
displacements of the rf electrodes on the effective potential was not known. A
tilt of one set of rf electrodes by a tenth of a degree can therefore not be
excluded for our trap. Such small tilt angles already come close to the
mechanical tolerances for the assembly of a 22-pole trap in the presently used
design.  This shows the need to significantly improve the precision in the rf
electrode geometry when a potential with pocket depths in the $\mu$eV range is
desired.

In searching for an explanation for the observed ten minima, we have extended
the electric field simulations to multipoles of different order $n$. These
calculations have shown that the number of minima observed in the effective
potential of a distorted multipole ion trap is directly connected to the
multipole order as $N_{\rm minima} = n-1$. Besides the above discussed tilt of
one set of electrodes, other distortions of the ion trap, such as a parallel
displacement of one set of rf electrodes, also introduces $n-1$ minima. The
number of minima therefore seems to be a general consequence of breaking the
symmetry of a multipole ion trap. We expect the minima to be related to the
points in space where the superimposed time-dependent electric field of the
individual multipole electrodes cancels, because according to
Eq.\ (\ref{effective_potential}) these are the global minima of the effective
potential. For a perfectly symmetric multipole trap cancellation is expected
only in the trap center, but for a distorted symmetry several such points are
found at larger radii.

\section{\label{sec:conclusion}Conclusion}
In this article we report on a method to directly measure the column density
distribution of ions in a 22-pole ion trap using photodetachment of stored
OH$^-$. The two-dimensional tomography scans yield the effective potential
averaged over the length of the trap. The measurements quantitatively confirm
the overall validity of the effective potential of a 22-pole ion trap, which
scales as $r^{20}$. For large rf amplitudes, however, new features in the
potential have been observed in the form of ten almost equally spaced
potential minima. These minima arise from the breaking of the 22-fold symmetry
of the trap. They can be quantitatively explained by a slight tilt of half of
the multipole rf electrodes within their mechanical tolerances. Also for other
multipole ion traps $n-1$ minima in the effective trapping potential have been
found as a consequence of a broken symmetry of the trap.

This observation, which has become possible due to the high sensitivity of our
photodetachment tomography scans, has implications for other spectroscopic
experiments in 22-pole ion traps. In particular at cryogenic temperatures,
trapped ions will reside predominantly in the ten pockets of the
potential. Correspondingly, the ion density along the symmetry axis of the
trap would become very small and only a small spectroscopic signal would be
detected for a laser beam pointing along the trap axis. Further studies are
needed to find out if ions that reside in the ten pockets may be subject to
enhanced radiofrequency heating, similar to the influence of the static end
electrode potential, which can increase the translational temperature by a few
Kelvin \cite{asvany2009:ijm}. Generally, it is therefore advisable to operate
the ion trap at rf amplitudes far below 100\,V to significantly suppress the
pockets. To overcome the pockets a significantly enhanced precision in the
manufacturing and assembly of 22-pole traps is required. An interesting
alternative for precisely controllable multipole ion traps are planar,
chip-based traps \cite{debatin2008:pra}.

We thank Ferdinand Schmidt-Kaler for helpful discussions. This work is
supported by the Deutsche Forschungsgemeinschaft under contract No. WE
2592/2-1. P.H. acknowledges support by the Alexander von Humboldt foundation.
K.S. acknowledges support by the European commission within EMALI (Contract
No. MRTN-CT-2006-035369) and the Landesstiftung Baden-W{\"u}rttemberg in the
framework ’atomics’ (Contract No. PN 63.14) and the ’Eliteprogramm
Postdoktorandinnen und Postdoktoranden’.

\clearpage

\section*{References}

\providecommand{\newblock}{}

\newpage
\newpage

\begin{thebibliography}{10}
\expandafter\ifx\csname url\endcsname\relax
  \def\url#1{{\tt #1}}\fi
\expandafter\ifx\csname urlprefix\endcsname\relax\def\urlprefix{URL }\fi
\providecommand{\eprint}[2][]{\url{#2}}

\bibitem{gerlich1992:adv}
Gerlich D 1992 {\em Adv. Chem. Phys.\/} {\bf 82} 1

\bibitem{gerlich1995:ps}
Gerlich D 1995 {\em Phys. Scripta\/} {\bf T59} 256

\bibitem{Schlemmer1999Laser}
Schlemmer S, Kuhn T, Lescop E and Gerlich D 1999 {\em Int. J. Mass Spectrom.\/}
  {\bf 185-187} 589

\bibitem{schlemmer2002Laser}
Schlemmer S, Lescop E, von Richthofen J, Gerlich D and Smith M~A 2002 {\em J.
  Chem. Phys\/} {\bf 117} 2068

\bibitem{mikosch2004:jcp}
Mikosch J, Kreckel H, Wester R, Plasil R, Glosik J, Gerlich D, Schwalm D and
  Wolf A 2004 {\em J. Chem. Phys.\/} {\bf 121} 11030

\bibitem{asvany2005:sci}
Asvany O, Kumar P, Redlich B, Hegemann I, Schlemmer S and Marx D 2005 {\em
  Science\/} {\bf 309} 1219--1222

\bibitem{Glosik06}
Glos\'{i}k J, Hlavenka P, Pla\v{s}il R, Windisch F, Gerlich D, Wolf A and
  Kreckel H 2006 {\em Phil. Trans. R. Soc. A\/} {\bf 364} 2931

\bibitem{boyarkin2006:jacs}
Boyarkin O~V, Mercier S~R, Kamariotis A and Rizzo T~R 2006 {\em J. Am. Chem.
  Soc.\/} {\bf 128} 2816

\bibitem{dzhonson2006:jms}
Dzhonson A, Gerlich D, Bieske E~J and Maier J~P 2006 {\em J. Mol. Struct.\/}
  {\bf 795} 93--97

\bibitem{kreckel2008:jcp}
Kreckel H, Bing D, Reinhardt S, Petrignani A, Berg M and Wolf A 2008 {\em J.
  Chem. Phys.\/} {\bf 129} 164312

\bibitem{asvany2004:cp}
Asvany O, Savic I, Schlemmer S and Gerlich D 2004 {\em Chem. Phys.\/} {\bf 298}
  97--105

\bibitem{asvany2004:apj}
Asvany O, Schlemmer S and Gerlich D 2004 {\em Astroph. J.\/} {\bf 617} 685--692

\bibitem{otto2008:prl}
Otto R, Mikosch J, Trippel S, Weidem{\"u}ller M and Wester R 2008 {\em Phys.
  Rev. Lett.\/} {\bf 101} 063201

\bibitem{wester2009:jpb}
Wester R 2009 {\em J. Phys. B\/} {\bf submitted}

\bibitem{dehmelt1967:adv}
Dehmelt H~G 1967 {\em Adv. At. Mol. Phys.\/} {\bf 3} 53

\bibitem{schubert1989:josa}
Schubert M, Siemers I and Blatt R 1989 {\em J. Opt. Soc. Am. B\/} {\bf 6} 2159

\bibitem{walz94}
Walz J, Siemers I, Schubert M, Neuhauser W, Blatt R and Teloy E 1994 {\em Phys.
  Rev. A\/} {\bf 50} 4122

\bibitem{mikosch2008:pra}
Mikosch J, Fr{\"u}hling U, Trippel S, Otto R, Hlavenka P, Schwalm D,
  Weidem{\"u}ller M and Wester R 2008 {\em Phys. Rev. A\/} {\bf 78} 023402

\bibitem{trippel2006:prl}
Trippel S, Mikosch J, Berhane R, Otto R, Weidem{\"u}ller M and Wester R 2006
  {\em Phys. Rev. Lett.\/} {\bf 97} 193003

\bibitem{hlavenka2009:acc}
Hlavenka P, Otto R, Trippel S, Mikosch J, Weidem{\"u}ller M and Wester R 2009
  {\em J. Chem. Phys.\/} {\bf 130} 061105

\bibitem{asvany2009:ijm}
Asvany O and Schlemmer S 2009 {\em Int. J. Mass. Spectrom.\/} {\bf 279} 147

\bibitem{greengard1988}
Greengard L 1988 {\em The Rapid Evaluation of Potential Fields in Particle
  Systems\/} (M.I.T. Press)

\bibitem{nabors1994:jsc}
Nabors K, Korsmeyer F~T, Leighton F~T and White J 1994 {\em SIAM J. Sci. Stat.
  Comp.\/} {\bf 15} 713

\bibitem{debatin2008:pra}
Debatin M, Kr{\"o}ner M, Mikosch J, Trippel S, Morrison N, Reetz-Lamour M,
  Woias P, Wester R and Weidem{\"u}ller M 2008 {\em Phys. Rev. A\/} {\bf 77}
  033422

\end{thebibliography}
\end{document}